# Mobile Agents for Distance Evaluation Procedures

***PENTIUC Stefan Gheorghe, GIZA Felicia, SCHIPOR Ovidiu Andrei***

*"Ştefan cel Mare" University, Suceava, Romania,
pentiuc@eed.usv.ro, felicia@eed.usv.ro, schipor@eed.usv.ro*

**Abstract:**
*The growth of Internet has led to new approaches for distance education. The main mechanisms involved in this process are distance learning and distance evaluation. Distance learning have multiple forms, from browsing and finding information when needed or collecting information onto ready-to-use packages, to interactive content where the learner is able to affect and control the content in some way. From the teacher's point of view, evaluation of learning aims to determine whether the students have achieved the goals set for a particular topic or course. Typically this is tested somehow and based on the test results, evaluation is performed. This kind of method is always restricted to the test, however. Many times the students may actually learn something completely different from those that are tested. Furthermore, testing does not necessarily take the students' individual needs into account. However, when properly designed, the testing method is a practical and even fairly reliable way of evaluating learning.*

**1. Traditional techniques for distance evaluation**

Distance evaluation can have various types, but they all fall into one of the two approaches [Marian et al., 2004] considered below:
- a pull scenario, the self-assessment approach, when the learner is initiating the evaluation to get his current level of knowledge.
- a push scenario, for example an exam, initiated by the teaching authority enforcing one particular test to assess all student's knowledge level. The test results shall be stored by the system [Rahkila, 2006].

Most of the present day Internet based evaluations are web centric and employ the client-server paradigm. The process of evaluation depends upon a local database of the questions (typically on a local area network) and does not scale well for remote testing.

In web based testing [Jamwal et Iyer, 2003] on the client side the students download a questionnaire as a web page and submit the answers back to the server. The server evaluates the answers and returns the results to the client. Existing Internet evaluation mechanisms, such as web based testing, rely principally on the client-server model, described above. Such mechanisms usually do not scale well and also do not fully support features like: evaluation of subjective questions, delivery of dynamic content, and off-line examinations. These features are extremely desirable for distance evaluation and there is a need for alternate ways of designing such applications.

Computer-based education makes it possible to automates, at least partly, the testing process. If the test questions are limited to simple, unambiguous answers, it is possible to directly evaluate the results. This kind of approach is fast and gives the students immediate feedback. However, the applicability of this kind of tests is somewhat questionable since in most cases, the test would be too limited with respect to the course or topic.

Another approach is to use computers for gathering the results and allow the teacher to evaluate the answers. This method allows wide selection of question types to be used in the test, but it only helps in getting the answers in a compact way. Otherwise, the method is similar to standard exams.

Somewhere in between is a "semi-automatic" approach, where software is used for gathering the results as well as partially evaluating them. This kind of approach is nowadays perhaps the most popular in the field of computer assisted testing and exams.

### 2. Distance evaluation Application scenario

We consider an examination scenario where a group of students are evaluated concurrently. A typical large-scale examination process involves the following stages:

(i) Setting and initialization of the inference engine for the intelligent evaluation. The exam should consist in a number of tests evaluated by an expert system that has the possibility of setting the questions and the path followed by different student based on their previous answers. The particular test and the inference engine should be carried by a mobile agent to the student computer.

(ii) Distribution of test to the enrolled students. This stage involves sending the particular test and the inference engine as a mobile agent to student's computer. Each student computer must have installed compatible agent platforms. After the designated examination duration or when the student finishes, each mobile agent returns to the home platform with the student's results.

(iii) Evaluation, compilation and publication of the results. This stage involves compilation of the results and their publication. When a mobile agent reaches the server, the results are then compiled and published.

### 3. Mobile agents for distance evaluation

Over the past few years, the Mobile Agent paradigm has emerged as a new mechanism for structuring distributed applications. It promises to alleviate many of the shortcomings of the client-server approach.

Mobile agent is an autonomous piece of software that can migrate between the various nodes of the network and can perform computations on behalf of the user. Some of the benefits provided by mobile agents include reduction in network load, overcoming network latency and disconnected operations.

Mobile agents being autonomous and dynamic entities offer many advantages over traditional design methodologies in distance evaluation [Jamwal et Iyer, 2002]:
- Mobile agents can package up a conversation and ship it to a destination host, where the interactions can take place locally, so it can avoid networking delays, improve the bandwidth limitation and adapt the test on the student performance.

- Mobile agent's technology naturally supports mobile devices and mobile users; it can promote evaluation environment's usability.
- Mobile agent systems are generally computer and network independent, they support distributed systems and resources sharing. So it can solve the heterogeneous problem of computers and networks on which the system is built.
- Mobile agents offer support for push. There are cases where pushing information to the users is a better alternative than the users pulling the information from the servers. For example, such a need may arise when some run-time notices are to be communicated to the students.
- Mobile agents are suited to work off-line or intermittent connectivity to Internet as long as all the computation is locally.

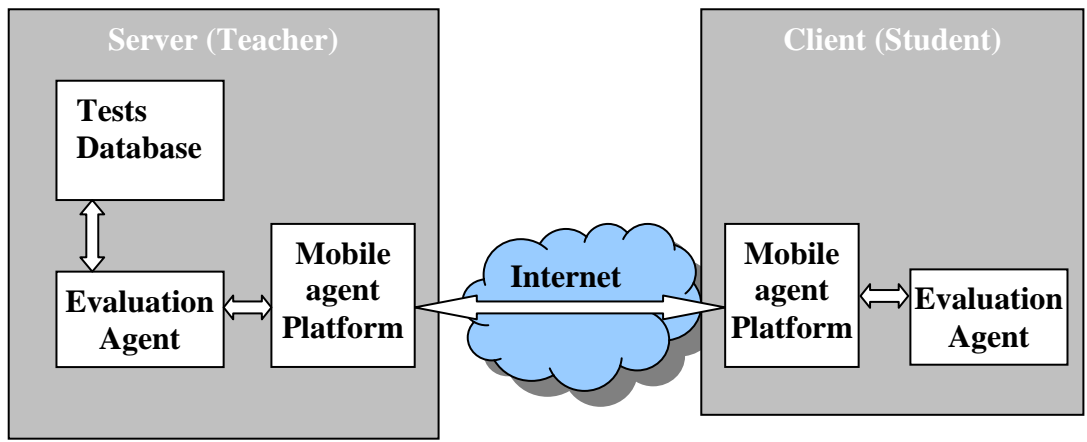

Figure 1.Architecture of evaluation process

Figure 1 describes the implementation of mobile agent's technology in the process of evaluation. In high-level view, basically mobile agent (Evaluation Agent) interacts with server agent to pick up data and inference engine which rely on particular test. Then mobile agent moves to host (or client) side. At host side, mobile agent performs all the processes needed for the evaluation of the student. After mobile agent finishes all behaviors in host side, it gathers all information it need, and return to server side.

The system needs an agent platform that supports mobility to be installed on the teacher computer as well on all participating student's computers.

The Tests Databases contains a lot of tests for different modules from the teacher's course. To enable automatic evaluation, the questions should have exact answers, in the form of different choices or writing a short number of words.

The entire process of evaluation is coordinated by a Teacher Agent, which is a stationary agent that lives on the server machine and is responsible with handling compulsory examinations, generating corresponding evaluation engines and gathering the result tests after de exam.

For the evaluation itself we considered an Evaluation Agent. The Evaluation Agent is a mobile agent that encapsulates all the necessary data and knowledge to perform an assessment, migrates on the client (Student) machine and is able to perform the evaluation. Evaluation Agent is loaded with an evaluation engine containing the complete test (particular questions based on previous answer, answer options, correct answer) and the assessment procedure. When the evaluations process ends, he goes back on the server and delivers the results for centralization.

### 4. Advantages of this approach

The mobile agent technology can overcome some limitations of the well known server-client model [Anghel, 2003]:

1. Scalability issues for many users – mobile agent based design enables efficient addition (or even deletion) of nodes participating in an application. In our system, new student nodes can be added to the system without affecting the performance of the system. Since each of these nodes is autonomous, its addition does not unduly load any single component of the system.

2. bandwidth / latency – transferring all the evaluation engine information in one step (by using a mobile agent) reduces the overall network communications overheads, and also allows the user to access all required information instantly. More over, the evaluation of the test can be performed directly on the client's computer. In this way the server is freed from the task of grading students (which can be quite a consuming operation)

3. Flexible structure - A change in system architecture will just involve supplying the new installation rules and components to an Install Agent which is send to apply the changes to each host from the system. .

4. Support for push and pull – Mobile agents enable an application to exploit both push and pull models of content delivery. A need for push model exists in cases like distribution of run-time notices that must be communicated to the students